\documentclass{article}

\PassOptionsToPackage{numbers}{natbib}

\usepackage[preprint]{neurips_2023}

\usepackage[utf8]{inputenc} %
\usepackage[T1]{fontenc}    %
\usepackage{hyperref}       %
\usepackage{url}            %
\usepackage{booktabs}       %
\usepackage{amsfonts}       %
\usepackage{nicefrac}       %
\usepackage{microtype}      %
\usepackage{xcolor}         %
\usepackage{graphicx}
\usepackage{booktabs}
\usepackage[normalem]{ulem}
\useunder{\uline}{\ul}{}
\usepackage{xcolor}
\usepackage{hyperref}
\usepackage{subcaption}
\usepackage{multirow}
\usepackage{amssymb}
\usepackage{amsmath}

\title{Effective Matching of Patients to Clinical Trials using Entity Extraction and Neural Re-ranking}

\newcommand{\absdiv}[1]{%
  \par\addvspace{.5\baselineskip}%
  \noindent\textbf{#1}\quad\ignorespaces
}

\author{%
Wojciech Kusa$^{1\dagger}$ \quad Óscar E. Mendoza$^{2\dagger}$ \\
\textbf{Petr Knoth}$^3$ \quad \textbf{Gabriella Pasi}$^2$ \quad \textbf{Allan Hanbury}$^1$ \\
$^1$TU Wien \quad $^2$University Milano-Bicocca \quad $^3$The Open University\\
\texttt{\{wojciech.kusa,allan.hanbury\}@tuwien.ac.at}\\
\texttt{\{oscar.espitiamendoza,gabriella.pasi\}@unimib.it}\\
\texttt{petr.knoth@open.ac.uk} \\
$^{\dagger}$Equal Contribution
}

\begin{document}

\maketitle

\begin{abstract}
\absdiv{Introduction}
Clinical trials (CTs) often fail due to inadequate patient recruitment.
Finding eligible patients involves comparing the patient's information with the CT eligibility criteria.
Automated patient matching offers the promise of improving the process, yet the main difficulties of CT retrieval lie in the semantic complexity of matching unstructured patient descriptions with semi-structured, multi-field CT documents and in capturing the meaning of negation coming from the eligibility criteria. 
\absdiv{Objectives}
This paper tackles the challenges of CT retrieval by presenting an approach that addresses the patient-to-trials paradigm.
Our approach involves two key components in a pipeline-based model: (i) a data enrichment technique for enhancing both queries and documents during the first retrieval stage, and (ii) a novel re-ranking schema that uses a Transformer network in a setup adapted to this task by leveraging the structure of the CT documents.
\absdiv{Methods}
We use named entity recognition and negation detection in both patient description and the eligibility section of CTs. 
We further classify patient descriptions and CT eligibility criteria into current, past, and family medical conditions. 
This extracted information is used to boost the importance of disease and drug mentions in both query and index for lexical retrieval. 
Furthermore, we propose a two-step training schema for the Transformer network used to re-rank the results from the lexical retrieval. 
The first step focuses on matching patient information with the descriptive sections of trials, while the second step aims to determine eligibility by matching patient information with the criteria section.
\absdiv{Results}
Our findings indicate that the inclusion criteria section of the CT has a great influence on the relevance score in lexical models, and that the enrichment techniques for queries and documents improve the retrieval of relevant trials.
The re-ranking strategy, based on our training schema, consistently enhances CT retrieval and shows improved performance by 15\% in terms of precision at retrieving eligible trials.
\absdiv{Conclusion}
The results of our experiments suggest the benefit of making use of extracted entities.
Moreover, our proposed re-ranking schema shows promising effectiveness compared to larger neural models, even with limited training data.
These findings offer valuable insights for improving methods for retrieval of clinical documents.
\end{abstract}

\section{Introduction}

Clinical trials (CTs) are crucial to the progress of medical science, specifically in developing new treatments, drugs, or medical devices~\cite{pressler2012computational}.
Awareness and access to these studies are still challenging both for patients and physicians, making the recruitment of patients a significant obstacle to the success of trials~\cite{ni2015automated,pressler2012computational}.

Even if a sufficient number of patients is found, the recruitment process requires screening the patients for eligibility, which is a labour-intensive task~\cite{embi2008physicians}. 
Automated identification of eligible participants not only promises great benefits for translational science~\cite{ni2015automated} but also aids patients by allowing them to be included in specific trials~\cite{koopman2016test}. 

In recent years, several initiatives have been proposed to build automatic systems for matching patients to CTs~\cite{shivade2015textual,koopman2016test,roberts2017overview,roberts2021overview}. 
The task has been defined as an information retrieval problem under the patient-to-trials evaluation paradigm~\cite{roberts2019overview}. 
Here, the query is constituted by patient-related information, either in the form of electronic health records (EHRs) or ad-hoc queries, and the documents are the CTs~\cite{koopman2016test}.

This retrieval task involves the semantic complexity of matching the patients' information with heterogenous, multi-fielded CT documents~\cite{rybinski2020clinical}. 
In addition to this, the eligibility criteria often use complex language structures (e.g. concepts can be negated twice) and medical jargon given in either semi-structured or unstructured ways~\cite{dasgupta2021automatic}.

To date, the existing approaches have revealed a significant lack of balance between efficiency and effectiveness. While pipeline-based models showcase promising performance, the substantial model sizes required to achieve competitive results raise concerns regarding costly deployment and limitations on reproducibility.
This work presents a system for CT matching that uses data enrichment techniques for supporting CTs search with probabilistic lexical model as fist retriever, and a re-ranking setup with a Transformer network with a moderate size. 

On the one hand, we develop a data enrichment process for both queries and documents. 
It consists of entity recognition and negation detection modules, applied to both the patient's description and the eligibility section of CTs. 
The data enrichment process also provides the classification of both patient's descriptions and CT eligibility criteria into current, past and family-medical conditions.
The extracted information boosts the importance of affirmative and negative mentions of diseases and drugs for both the documents and queries in the traditional retrieval scenario. 

On the other hand, we define a training strategy for re-ranking trials using a pre-trained language model in a two-step schema that leverages the structure of CT by considering not only the traditional topical relevance objective but also the eligibility criteria. 
Taking the result from our first stage retrieval process, we then match patient's information with descriptive sections of the trials for re-ranking based on  topical relevance. 
Later, we further train this model by matching patient data with trial eligibility criteria in an attempt to discriminate documents as eligible or excluded.

We evaluate our work on the TREC Clinical Trials track 2021 and 2022 collections, showing that our methods improve finding relevant trials. 
More specifically, our contributions are as follows:

\begin{enumerate}
    \item We evaluate the utility of individual sections of CT text on the performance of the lexical retrieval system showing that the inclusion criteria section alone contributes the most to the effectiveness of the search system.
    \item We introduce a new query and document enrichment model that uses information extraction modules to handle challenges posed by unstructured EHR descriptions and eligibility criteria sections of CTs.
    The additional data explicitly highlight sections of patients' medical history and establish a novel way of handling a negation from the eligibility criteria.
    Rather than relying on dictionaries to find these entities, we use neural network-based information extraction models.
    \item We propose and develop an effective re-ranking setup adapted to CT retrieval considering different learning objectives for training.
    We evaluate its quality both on the general, pre-trained BERT model, as well as biomedical domain-focused versions.
\end{enumerate}

\section{Background}

This section describes previous work on CT matching with various paradigms, approaches to extract information from clinical data and from patient-related information, and neural re-ranking for CT retrieval.

\subsection{Clinical trials matching}

The TREC Clinical Trials track concerns the task of matching single patients to clinical trials.
Other tasks concerning CT matching mentioned in the literature are cohort-based retrieval~\cite{koopman2021cohort} and trial-to-trial retrieval~\cite{wang2022trial2vec}.

In the context of the TREC CT track, patient-related information is  written as free-text, whereas the document collection consists of a snapshot of ClinicalTrials.gov database\footnote{\url{https://clinicaltrials.gov}}.
Each clinical trial contains multiple fields, including two titles (brief and official one), condition, summary, detailed description, and eligibility criteria. 
The content of these fields can range from structured (e.g. gender and age of eligible patients) through semi-structured (e.g. eligibility criteria section) to unstructured (e.g. description and summary).
The eligibility criteria field contains inclusion and exclusion criteria, a core aspect of the CT matching task.
Trials were judged using a graded relevance scale of three points: $0$ if the patient is not relevant to the CT, $1$ if the patient is topically relevant but excluded based on the eligibility criteria, and $2$ when the patient fulfils the eligibility criteria.

The TREC CTs track differs from previous medical TREC tracks in several aspects. 
TREC Precision Medicine 2017--2020~\cite{roberts2017overview} is concerned with matching CTs to a patient summary consisting of only the patient's disease, relevant genetic variants, and basic demographic information. 
On the other hand, TREC CT topics consist of an unstructured patient summary.
TREC Clinical Decision Support 2014--2016~\cite{roberts2015overview} used topics written similarly (free-text patient descriptions), but the task was to match patients to PubMed publications, instead of CT documents.
Moreover, none of the previous tracks used a graded relevance scale focused on eligibility.

\begin{figure}[t]
    \centering
    \includegraphics[width=1\textwidth]{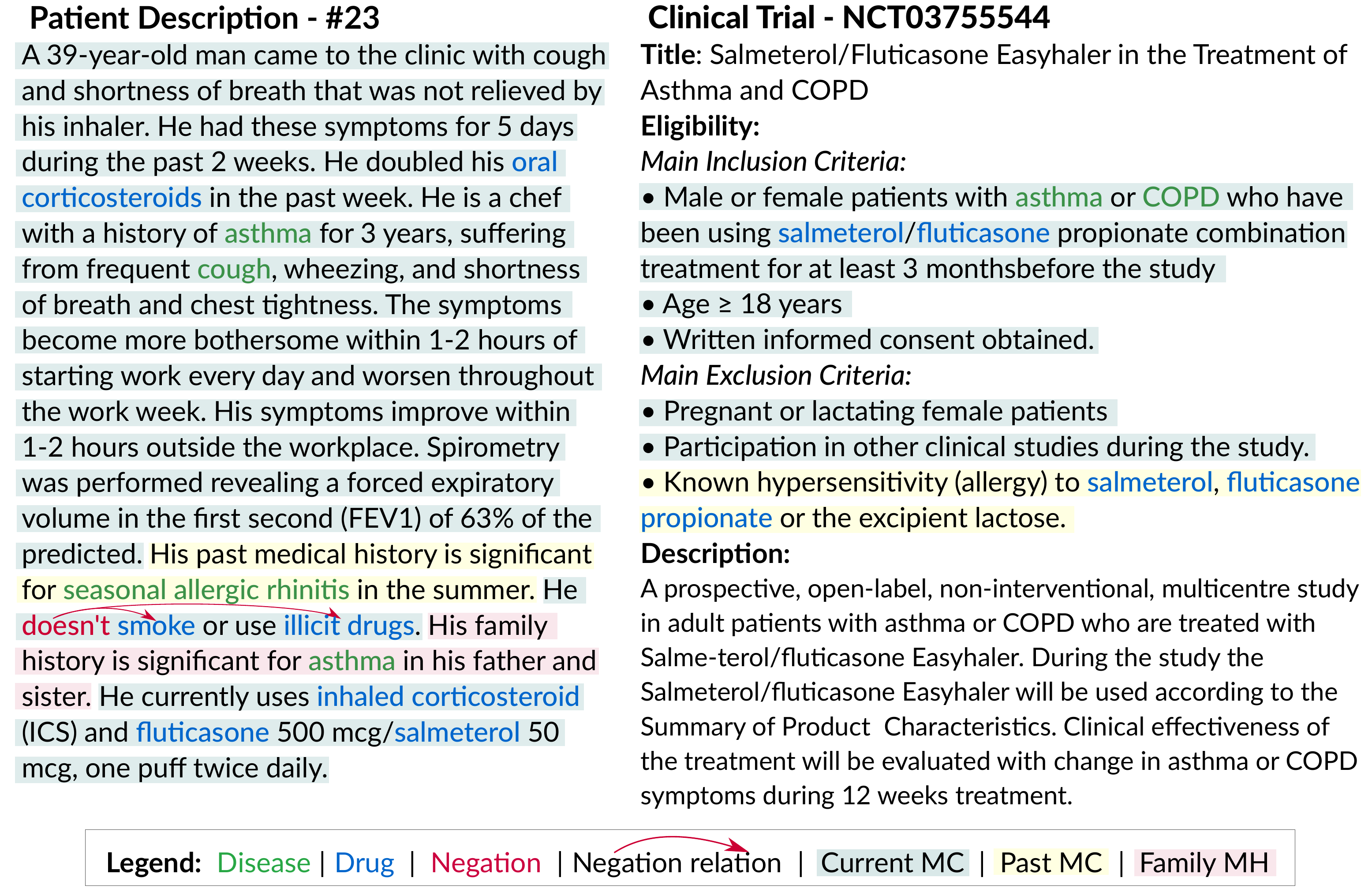}
    \caption{An example of a clinical trial and a description of a patient eligible for this trial. Highlighted items are described in detail in Section \ref{sec:meth}. Example adapted from~\citet{pradeep2022neural}.}
    \label{fig:example_trial_patient}
\end{figure}

Figure \ref{fig:example_trial_patient} provides an example of a patient's EHR description and of the sections from a relevant CT.
Using a bag-of-words approach to tackle the patient-to-trial matching problem may pose difficulties as both the patient's description and the CTs contain many irrelevant terms, thereby introducing noise.
Moreover, both can contain negated key terms (for instance, the exclusion criteria), the handling of which is essential for deciding eligibility but may not be trivial even when using n-grams or neural network-based models~\cite{grivas2020not,sykes2021comparison}. 
Additionally, the sections of queries and documents may have different importance because of their time dependency (i.e., past or present conditions) and because they can refer to either patients or patients' family medical history.
One can try to overcome these issues by structuring both the query and documents, and extracting relevant items first.

Previous work attempted to solve a CT matching task using various lexical and neural models.
\citet{leveling2017patient} annotated a corpus with terms from medical dictionaries and with negations for retrieving trials for the TREC Precision Medicine track. %
A large number of systems reported in the TREC CT used variants of the Okapi BM25 model~\cite{jones2000probabilistic} or the Divergence from Randomness (DFR) model~\cite{amati2002probabilistic} that has demonstrated potential in the biomedical information retrieval field.

\subsection{Information extraction from clinical data}

Information extraction from clinical data has been an active area of research in recent years. 
Previous work has focused on automatic extraction of eligibility criteria from clinical trial protocols. 
For instance,~\citet{dasgupta-etal-2020-extracting} presented a method for identifying and segmenting eligibility criteria into five semantic categories, including demographic information, health status, treatment history, laboratory test reports, and lifestyle.
The EliIE system~\cite{10.1093/jamia/ocx019} was proposed for converting free-text eligibility criteria for clinical research into a structured and formalised format using a 4-step process including entity and attribute recognition, negation detection, relation extraction, normalization of concepts and output structuring.

Other studies aimed to extract information from patients' health records. 
The development and uptake of NLP methods for processing free-text clinical notes has been growing in recent years. 
A systematic review of the literature by~\citet{info:doi/10.2196/12239} showed that there is a significant increase in the use of machine learning methods for NLP in clinical notes related to chronic diseases, and that deep learning is an emerging methodology.
The ConText algorithm aims to determine whether conditions mentioned in clinical reports are negated, hypothetical, historical, or experienced by someone other than the patient~\cite{harkema2009context}.
The n2c2 n2c2/OHNL 2019 shared task~\cite{info:doi/10.2196/24008} focused on extracting family history information from clinical notes.
\citet{garcelon2017improving} utilised heuristics to detect medical history and negated terms in patients' records.

Despite these efforts, there has been a lack of approaches that integrate information extraction techniques to enhance both query and document representation.
Specifically, there is a lack of methods that effectively combine the extracted terms to determine eligibility ranking.
This presents an opportunity for further exploration in the field.

\subsection{Neural approaches for CT}

Several approaches using Transformer-based architectures and pre-trained models, such as BERT~\cite{bert}, have achieved state-of-the-art effectiveness in some of the biomedical information processing applications. 
In CT retrieval, there have been multiple attempts to use BERT embeddings in both dual-encoder and cross-encoder retrieval setups with different pre-trained models such as BioBERT or ClinicalBERT~\cite{jinalibaba,Rybiski2022CSIROmedTR,rybinski2020clinical}. 
These results correspond to implementations of methods applied to traditional ad-hoc retrieval tasks and have not outperformed multiple experiments under traditional retrieval models~\cite{roberts2021overview,roberts2022overview}. 
On the other hand,~\citet{pradeep2022neural} proposed a multi-stage neural ranking system for the CTs matching problem, relying on T5-based models, currently with state-of-the-art results in multiple retrieval tasks, including CT. 

According to the findings presented in TREC CT 2021~\cite{roberts2021overview}, T5-based models currently outperform smaller transformers models in CT retrieval. In this paper, we propose an effective training strategy that takes into account various aspects of clinical trial retrieval, including topical relevance and eligibility criteria, as separate learning objectives. We evaluate its quality both on the general, pre-trained BERT model, as well as biomedical domain-focused versions. Our approach results in a strong competitor to T5-based models with a much simpler architecture, as demonstrated by the official results reported in TREC CT 2022~\cite{roberts2022overview}. Specifically, our model performs second-best overall, outperformed only by the model proposed by~\citet{pradeep2022neural} in the best-performing category. These findings suggest that BERT-based models with our proposed training strategy can provide a viable alternative to T5-based models in clinical trial retrieval.

\section{Methodology} \label{sec:meth}

This section describes the steps for processing CTs' and patients' descriptions used as input to probabilistic lexical models.
We then define our two-stage neural re-ranking pipeline.

\subsection{Clinical trial processing and ranking} \label{sec:ct}

We parse the content of a clinical trial document to split it into specific sections.
The \emph{eligibility criteria} section contains a crucial component of the trial used to distinguish if a patient is eligible for a given trial.
Our CT processing is focused on making the eligibility criteria as fine-grained as possible so we can easily discriminate aspects referring to medical history and drugs. 
We start by using parser based on heuristics to split the eligibility criteria section of clinical trials into two lists containing inclusion and exclusion criteria, respectively.

We further classify each sentence from inclusion and exclusion as concerning one of the three sections: `\emph{current medical condition}', `\emph{past medical condition}' and `\emph{family medical history}'.
Our motivation is that admission notes (which the topics simulate), consist of several sections that do not have equal impact on the patients' relevance to the trial and may even be irrelevant to their eligibility.
Similarly, clinical trials can have different types of information stored in their eligibility section.

We then use a pre-trained entity extraction model together with an algorithm for determining negation to detect \emph{affirmative} and \emph{negative} drug and disease entities in both inclusion and exclusion sections.
In the next step, we remove double negations coming from negated exclusion criteria. 
For every entity in the exclusion criteria, we swap their modifier (from affirmative to negative and vice versa). 
For instance, the exclusion criterion `Patients who are \emph{not smoking}' becomes the inclusion criterion `Patients who are \emph{smoking}'. 
This step may not always be correct; nevertheless, we found it to be a good approximation, allowing us to collapse these two sections into one.

The final result is a single list of extracted entities, classified by their section and modifier.
All extracted keywords from a document $D_i$ can be described by the set $K_{D_{i}} = \{A_i^{cmc}, A_i^{pmc}, A_i^{fmh}, N_i^{cmc}, N_i^{pmc}, N_i^{fmh}\}$, where $A$ stands for a list with affirmative entities, $N$ for negative entities, and $cmc$, $pmc$ and $fmh$ for current medical conditions, past medical conditions, and family medical history, respectively.

We can then enrich the CT documents representation by expanding them with the extracted keywords.
However, in order to preserve the semantic information about each extracted entity (section and modifier information), we use prefixing with special tokens.
Furthermore, as many of these entities are multi-word expressions, we concatenate the tokens using the underscore character~`\_' to create a single token.
Specifically, we create new tokens by adding them the prefixes `$cmc$', `$pmc$' and `$fmh$' for each respective section and additionally `$no$' when an entity is negated (e.g. $N_i^{pmc} = [ $`$myasthenia$ $gravis$'$, $`$shortness$ $of$ $breath$'$]$ becomes $[$`$pmc\_no\_myasthenia\_gravis$'$,$\\ `$pmc\_no\_shortness\_of\_breath$'$]$).
We append the new tokens to the pre-processed document and use the enriched document to create an index for the lexical retrieval models.

\subsection{Patient description processing} \label{sec:patient}

As we are essentially aiming to match the patient to the CT criteria, we follow a similar approach to enrich the input query.
A patient's description is split into the sections of current medical conditions, past medical conditions, and family medical history. 
As for the trials, we run an entity and negation detection algorithm for each section.
Extracted keywords for query $Q_j$ can be represented as $K_{Q_{j}} = \{A_j^{cmc}, A_j^{pmc}, A_j^{fmh}, N_j^{cmc},$ $ N_j^{pmc}, N_j^{fmh}\}$, where each element contains a list of extracted entities.
Finally, after tokenisation, the query for lexical models containing the original patient description is enriched by appending the extracted entities.

\subsection{Filtering} \label{sec:method-filtering}

Following approaches from previous work~\cite{leveling2017patient,kusa2021dossier,rybinski2020clinical}, we employ filtering on the age and gender to eliminate trials for which patients would be excluded based on the demographic criteria.
We parse the age and gender information from patient descriptions for all patients.
In clinical trials, this corresponds to `\emph{minimum\_age}', `\emph{maximum\_age}' and `\emph{gender}' fields.
In this step, we remove the trials for which the patient is ineligible based on these two criteria.

Furthermore, we try rule-based parsing to extract information about smoking and alcohol consumption from both patients and clinical trials. 
Similarly to the demographic filters, we use this information to filter out ineligible patients.

\subsection{Re-ranking}
\label{sec:reranking}

Taking advantage of the structure of the documents and the topic processing discussed in Sections \ref{sec:ct} and \ref{sec:patient}, respectively, we define a training schema with two objectives. Here, inspired by the notion of curriculum learning, the approach aimed at decomposing complex knowledge and designing a curriculum for learning concepts from simple to hard~\cite{zeng2022curriculum}, we follow the heuristic that the CT retrieval task can be decomposed into two sub-tasks. First, we set the retrieval objective, which simply relies on discriminating topical relevance (both eligible and excluded documents are relevant). Second, we set the objective of eligibility classification (only eligible documents are relevant). 

We use the pre-trained language model BERT~\cite{bert} with the standard approach known as cross-encoder neural ranking model. For fine-tuning, a linear combination layer is stacked atop the Transformer network, whose parameters are tuned with a ranking loss function. We use a pairwise loss function and train the model for re-ranking outputs from the process described in previous sections.

Thus, the model is trained for these two objectives consecutively, such that there are two instances of the same model that we optimise with the following loss:

\begin{equation}
    \mathcal{L}(q,d^+,d^-;\phi)=\text{max}(0,1-s(q,d^+;\phi)+s(q,d^-;\phi)),
\end{equation}
where $d^{(+)/(-)}$ denotes embedded-relevant or non-relevant documents to the topic representation $q$, $\phi$ represents the model's parameters with the final linear layer, and $s$ is the predicted score.

\begin{figure}[ht]
    \centering
    \includegraphics[width=1\textwidth]{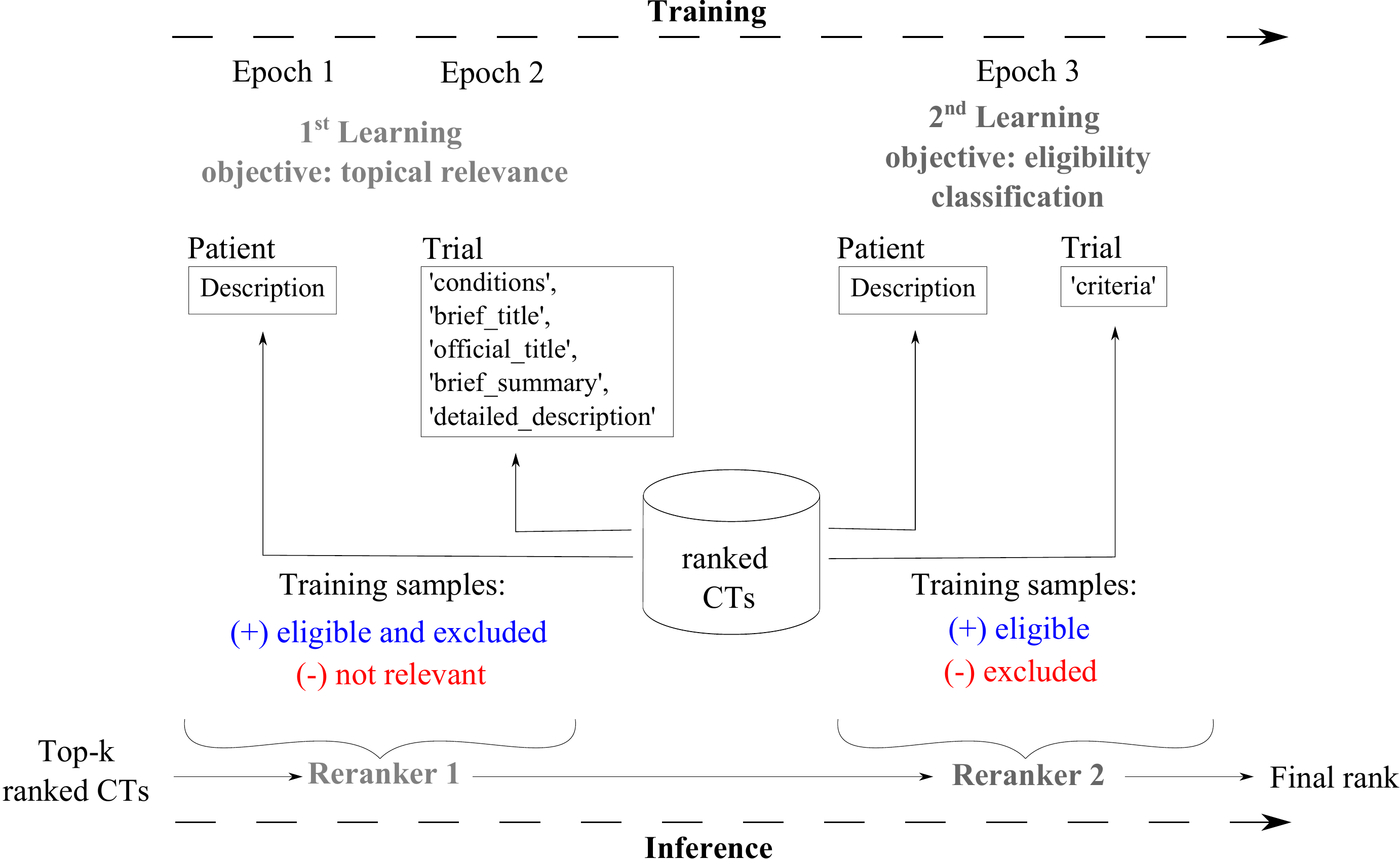}
    \caption{Neural re-ranking setup.}
    \label{fig:neural-reranking}
\end{figure}

As shown in Figure \ref{fig:neural-reranking}, we match patient information with descriptive sections of the trials for re-ranking based on topical relevance ($+d$ corresponds to sections of relevant trials). 
We consider the eligibility classification a harder task.
Moreover, we hypothesise that for this task, the model could benefit from the knowledge that it already has from the previous training. 
We further train this model by matching patients' information with criteria in an attempt to discriminate documents as eligible or excluded ($+d$ corresponds to trials categorised as eligible, and $-d$ corresponds to trials categorised as relevant but discarded). 

This process results in two different models. During inference time, we follow a similar schema: we take the BM25 rank and re-rank twice the top-k ranked trials using the two resulting models, respectively. When referring to this re-ranking procedure we call it \textbf{TCRR}: \emph{Topical and Criteria Re-Ranking}.

\section{Experiment setup}

This section details the datasets and baselines we have employed as well as the evaluation procedure. 
Additionally, we discuss the implementation of the methods described in the paper.

\subsection{Dataset}

The corpus released by TREC contain 375,580 clinical trials. In 2021, 75 topics (patient notes) were used, and 50 more were created in 2022. 
There are 35,832 relevance judgements in 2021 and 35,394 in 2022. 
More details of the datasets can be found in Table \ref{tab:trec-ct-stats} of Appendix \ref{app:datasets}. 
Clinical trial documents released by TREC are in \textsc{xml} format and consist of several sections.
In our experiments we consider the following sections: brief title, official title, description, summary, conditions and criteria.

For our experiments, we use the sets of topics as they where provided. 
For neural re-ranking, we present experiments using the topics from 2021 for training and from 2022 for testing and vice-versa. 
Additional splitting for training and development for neural models is described in Section~\ref{implementation}.

\subsection{Evaluation}

We follow the evaluation procedure from the TREC Clinical Trials track, which is the standard evaluation procedure for ad-hoc retrieval tasks. 
As the relevance assessment is given in a graded relevance scale (eligible, excluded, or not relevant), the performance of the models is measured using normalised discounted cumulative gain (nDCG). 
We present results as reported by TREC, using nDCG@5 and nDCG@10, Precision (P@10), and Reciprocal Rank (RR).

We treat unjudged documents as non-relevant, ensuring that our results are not biased towards models that retrieve a large number of unjudged documents.
Furthermore, we focus on Precision as the primary metric for optimising retrieval models.
Our goal is to identify eligible trials, and Precision provides strict feedback to achieve this aim.

\subsection{Baselines}
\label{baselines}
As discussed in Section \ref{sec:reranking}, for our custom re-ranking we train two different models: TopicalRR and CriteriaRR.
When used independently, we consider them as baselines: 

\paragraph{TopicalRR} The model trained for re-ranking based on the topical objective is  initialised with the weights of \emph{bert-base-uncased}\footnote{\url{https://huggingface.co/bert-base-uncased}} (as well as other two domain-specific trained models: BioBERT\footnote{\url{https://huggingface.co/seiya/oubiobert-base-uncased}} and Clinical-BERT\footnote{\url{https://huggingface.co/Tsubasaz/clinical-pubmed-bert-base-512}}). 

\paragraph{CriteriaRR} The model trained for re-ranking based on the eligibility criteria classification objective is initialised with the weights of the TopicalRR. We further train this model.

Additionally, we consider the following two neural models as baselines:

\paragraph{TraditionalRR} The cross-encoder we use to compare our proposed training procedure with the traditional training, we train the model from the same checkpoint \emph{bert-base-uncased}.

\paragraph{MonoBERT} One of the comparable models implemented from the TREC CT track. It is based on the cross-encoder architecture and trained on data drawn from the corpus in a weakly supervised fashion~\cite{roberts2022overview}.

\subsection{Implementation details}
\label{implementation}
We use the ScispaCy~\cite{neumann-etal-2019-scispacy} and medspaCy~\cite{eyre2021medspacy} to implement our entity extraction experiments.
We apply the spaCy NER model trained on the \textsc{bc5cdr} dataset to detect disease and drug mentions.

We have decided to use the ConText algorithm~\cite{harkema2009context} to determine whether extracted entities were negative or affirmative.
While more recent algorithms are available for identifying assertions in clinical text~\cite{van2021assertion}, we opted for the ConText algorithm due to its established track record and availability inside the ScispaCy library. 
Moreover, as our approach is focused on enriching not only 125 queries but also 375,000 clinical trial documents, an additional criterion for selecting the ConText model was its scalability.

Text is lowercased, and tokenised with the spaCy's \texttt{en\_core\_sci\_lg} model; punctuation and stopwords are removed.
As a main lexical retrieval model, we use the BM25+~\cite{trotman2014improvements} ``out-of-the-box'', i.e. without parameter optimisation, implemented in the Rank-BM25\footnote{\url{https://github.com/dorianbrown/rank\_bm25}} Python package.
Furthermore, for the first two experiments, we also test two other lexical models, namely TF-IDF~\cite{sparck1972statistical} and DFR model based on inverse document frequency with Bernoulli after-effect and H2 normalisation (In\_expB2)~\cite{amati2002probabilistic}, both implemented in the Terrier search engine\footnote{\url{http://terrier.org}}.

On the other hand, we use PyTorch Lightning~\cite{falcon_william_2019_7545285} and Transformers\footnote{\url{https://github.com/huggingface/transformers}} to implement the neural re-ranking pipeline. As discussed in Section~\ref{sec:reranking}, we train different models for re-ranking with different configurations (see Section \ref{baselines}).
The TopicalRR,  after splitting the datasets into train (60\%), development (10\%), and test (30\%) is trained on 8192 samples from the training set per epoch divided into batches of 16 samples. We further train this model on 1024 samples with batches of 16 to get to the CriteriaRR. Samples for these two models were selected as described in Section \ref{sec:reranking} and as shown in Figure \ref{fig:neural-reranking}. 
We pick positive samples only present in BM25 rankings as well as hard negatives from ranked-irrelevant or unlabeled documents.
We re-rank top-$50$ trials from the BM25 run\footnote{We ran experiments changing the cutoff from 20 to 100 with a step of 10 to find 50 as the optimal cutoff.}.
Finally, to compare our proposed training procedure with the traditional training of a cross-encoder, we train the TraditionalRR from the same checkpoint \emph{bert-base-uncased},  on 2048 samples, where relevant documents are only those categorised as eligible.

All neural models are trained for ten epochs, with early stopping based on Precision. Our training was performed on an Nvidia Quadro RTX 8000 GPU.

\section{Results}

We begin by assessing the effectiveness of using clinical trial sections.
Subsequently, we examine the influence of extracted entities and filtering techniques. 
Then, we conduct neural re-ranking experiments. 

\subsection{Clinical trials sections}

We first evaluate the utility of different sections of CTs.
We extracted inclusion and exclusion sections for 91\% of clinical trials.
For the remaining 9\% of trials, we assume that both criteria sections are empty.
We create several indexes and retrieval models with different combinations of sections as input features.
The results for the BM25+ model are presented in Table~\ref{tab:input_fields}.
The first eight rows represent results when only one section was used to create an index, whereas the remaining rows present runs conducted on the concatenations of selected sections.
Results for In\_expB2 and TF-IDF retrieval models are presented in Table \ref{tab:input_fields_other_models} of Appendix \ref{app:other_models}.

\begin{table}[!h]
\centering
\caption{Impact of CT sections on the performance of the retrieval model. 
The first group contains results using only a single section as a document representation, and the second group represents results using several concatenated sections.
All results use BM25+ as the retrieval algorithm. 
`Criteria' contains the value from the eligibility section, whereas `inclusion' and `exclusion' contains only the list of heuristically split section. 
{\ul Underlined} values indicate highest score within the group, \textbf{bold} values indicate highest score overall.
The identifier of each run is in the first column.}
\label{tab:input_fields}
\resizebox{\textwidth}{!}{%
\begin{tabular}{@{}l|l|cccc|cccc@{}}
\toprule
    &                                                 & \multicolumn{4}{c|}{\textbf{TREC CT 2021}}                                                     & \multicolumn{4}{c}{\textbf{TREC CT 2022}}                                                      \\ \midrule
    \textbf{\#}  
& \textbf{Input sections}                                  & \textbf{nDCG@5}  & \textbf{nDCG@10}   & \textbf{P@10} & \textbf{RR}                  & \textbf{nDCG@5}  & \textbf{nDCG@10}   & \textbf{P@10} & \textbf{RR}                  \\ \midrule
1.  & brief title                                    & .218                & .205                & .131                & .298                & .216                & .189                & .128                & .297                \\
2.  & official title                                 & .245                & .215                & .137                & .293                & .237                & .205                & .140                & .370                \\
3.  & description (\emph{desc.})                    & .354                & .317                & .195                & .408                & .324                & .277                & .168                & .381                \\
4.  & summary (\emph{sum.})                         & .332                & .315                & .192                & .376                & .346                & .305                & .220                & {\ul .480}          \\
5.  & conditions  (\emph{cond.})                   & .168                & .164                & .109                & .245                & .165                & .155                & .102                & .224                \\
6.  & inclusion                                       & {\ul.405}           & {\ul.391}           & {\ul.252}           & {\ul.478}           & {\ul .373}          & .337                & {\ul .230}          & .459                \\
7.  & exclusion                                       & .120                & .117                & .048                & .114                & .169                & .137                & .068                & .173                \\
8.  & criteria                                        & .397                & .367                & .199                & .411                & .363                & {\ul .338}          & .216                & .437                \\ \midrule
9.  & brief title + official title (\emph{tit.})  & .270                & .256                & .172                & .322                & .261                & .220                & .150                & .369                \\
10. & sum. + criteria + tit.                          & .470                & .445                & .255                & .467                & .450                & .427                & .292                & {\ul \textbf{.542}} \\
11. & desc. + criteria + tit.                         & .490                & .448                & .259                & .470                & .426                & .394                & .258                & .446                \\
12. & sum. + desc. + tit.                             & .402                & .386                & .243                & .443                & .414                & .381                & .272                & .491                \\
13. & sum. + desc. + tit. + cond. (\emph{all}) & .400                & .380                & .228                & .437                & .407                & .379                & .272                & .473                \\ 
14. & all + inclusion                                 & {\ul \textbf{.508}} & .462                & {\ul \textbf{.276}}                & {\ul \textbf{.505}} & .464                & {\ul \textbf{.437}} & {\ul \textbf{.312}} & .520                \\
15. & all + exclusion                                 & .398                & .367                & .203                & .395                & .386                & .363                & .238                & .451                \\
16. & all + criteria                                  & .491                & {\ul \textbf{.464}} & .272                & .492                & {\ul \textbf{.465}} & .426                & .290                & .506                \\ \bottomrule

\end{tabular}%
}
\end{table}

Among single section runs, the usage of the \emph{inclusion} field alone yields the highest results for Precision@10 and nDCG@5, both for 2021 and 2022 data.
Moreover, for 2021 topics, the \emph{inclusion} section also achieves the highest nDCG@10 and RR from all single topics, and it is on par with the run, which uses all sections except criteria combined (run 6 versus run 13).

Notably, for 2022, the \emph{summary} field achieves the highest RR among all single-field runs.
This is true for all three retrieval models.
This can be caused by having the first relevant trial more generic (i.e. covering broader or more common diseases) and relevant but not necessarily specific to the patient's conditions.
Figure \ref{fig:sections_2022_per_query} of Appendix \ref{app:topic_by_topic} shows a topic-by-topic comparison for RR and P@10 for the BM25+ model.
We can observe that there are still some topics for which the model using the \emph{inclusion} section achieves a higher RR score than the \emph{summary} field.

Concatenating more sections to create an index improves the on-average nDCG scores.
However, this does not always hold for the metrics that consider the distinction between eligible and ineligible (P@10 and RR).

The \emph{exclusion} section achieves the worst results from all single section runs (run 7), even when compared to runs using only the title of a clinical trial.
Moreover, simply adding the text from the \emph{exclusion} section for the bag-of-words approaches decreases the retrieval performance when compared to using the \emph{inclusion} section only (run 16 versus 14).
These outcomes motivate our subsequent experiments and document enrichment techniques described in Section~\ref{sec:ct}, where we try to structure the knowledge contained in the eligibility section to take advantage of the available data.

The results for In\_expB2 and TF-IDF (Table \ref{tab:input_fields_other_models} of Appendix \ref{app:other_models}) models follow a similar trend, with the differences for 2022 data even higher than for the BM25+ model.
This outcome shows that our findings can be generalised to other lexical models.

\subsection{Impact of extracted entities}

To determine the impact of the extracted entities, we selected the optimal configuration of input sections from the previous step, which used the summary, description, titles, conditions, and inclusion criteria (run 14).
We use these sections as a base document representation and enriched it with different combinations of extracted entities: c -- only current medical conditions, cf -- current and family medical history, cp -- current and past medical conditions, cfp -- current, family and past medical conditions.

\begin{table}[htbp]
\centering
\caption{Experimental results for runs with index and query expanded with extracted entities for BM25+ model. Letters describe usage of extracted affirmative and negative medical entities for (c) current conditions, (p) past conditions, and (f) family history. 
\textbf{Bold} values indicate highest score overall.
The identifier of each run is in the first column.
}
\label{tab:extraction-results}
\resizebox{\textwidth}{!}{%
\begin{tabular}{@{}l|r|cccc|cccc@{}}
\toprule
     & \multicolumn{1}{l|}{}      & \multicolumn{4}{c|}{\textbf{TREC CT 2021}}                               & \multicolumn{4}{c}{\textbf{TREC CT 2022}}                              \\ \midrule
     \textbf{\#}  
& \textbf{Model}    & \textbf{nDCG@5}  & \textbf{nDCG@10}   & \textbf{P@10} & \textbf{RR}             & \textbf{nDCG@5}  & \textbf{nDCG@10}   & \textbf{P@10} & \textbf{RR}            \\ \midrule
14.  & all + inclusion            & .508 & .462           & .276          & .505           & .464          & .437          & .312          & .520          \\ 
16.  & all +\ \ \  criteria       & .491          & .464           & .272          & .492           & .465          & .426          & .290          & .506          \\ \midrule
14a. & + c\ \ \ \ \               & .524          & .480           & .292          & .542           & .500 & .459 & .328 & \textbf{.528}          \\
14b. & + cf\ \ \ \                & .524          & \textbf{.481}           & \textbf{.293}          & .542           & .500           & .460          & \textbf{.330}          & \textbf{.528} \\
14c. & + cp\ \ \                  & .532          & .478           & .287 & \textbf{.555}           & .501          & .460          & .328          & .521          \\
14d. & + cfp\ \                   & \textbf{.532}          & .480 & .288 & \textbf{.555} & \textbf{.502}          & \textbf{.460}          & .328          & .521          \\ \bottomrule
\end{tabular}%
}
\end{table}

The results for the BM25+ model are presented in Table~\ref{tab:extraction-results}.
Using extracted items from patients positively impacts the final score.
The highest Precision scores are achieved with extracted affirmative and negated entities for the current and family medical history.
The low impact of past medical condition can be explained by an infrequent occurrence of this data in patient descriptions in the TREC dataset and the quality of the ConText algorithm.
Extracted entities contribute more positively to the measures where judgements distinguish between eligible and ineligible patients.
The best-performing model (14d) comprises all available extracted data (affirmative and negative entities for current, past and family medical history) to enrich the index.
This tells us that our proposed method can potentially improve the retrieval with complex negated sentences.
However, the relative performance gain is low, and a detailed analysis is needed to understand how it can be further improved.

An example of extracted entities is presented in Table~\ref{tab:example-extracted-patient}.
As can be seen, the performance of both entity extraction and section classification models generates both false positives and false negatives, which influences the final retrieval result.
Further fine-tuning on domain data could improve the quality.

\begin{table}[h]
\centering
\caption{Example entities extracted for Topic \#48 from TREC CT 2021.}
\label{tab:example-extracted-patient}
\resizebox{0.91\textwidth}{!}{
\begin{tabular}{p{1.13\linewidth}}
\toprule
\textbf{Topic \#48:} Fernandez is a 41 year man who is a professional soccer player. He came to the clinic with itchy foot. Physical exam revealed localized scaling and maceration between the third and fourth of his right toe. It became inflamed and sore, with mild fissuring. The dorsum and sole of the foot was unaffected. There is no pus or tearing in the affected area. He didn't use ant topical ointment on the lesion and has no positive history for any underlying disease such as DM. He smokes 15 cigarettes per day and drinks a beer per day. His family history is positive for hyperlipidemia in her mother and MI in her father. He is in relation with several partners and use condom during the intercourse. His physical exam and lab studies were normal otherwise. Tinea pedis infection confirmed as his diagnosis by the observation of segmented fungal hyphae during a microscopic KOH wet mount examination. \\
\bottomrule
\end{tabular}
}
\resizebox{0.5\textwidth}{!}{
\begin{tabular}{@{}ccc@{}}
\textbf{Section}            & \textbf{Entity}         & \textbf{Is negated} \\ \midrule
\multirow{6}{*}{Current MC} & itchy                 & ---                  \\
                            & sore                  & ---                  \\
                            & fissuring             & ---                  \\
                            & tearing               & $\checkmark$             \\
                            & Tinea pedis infection & ---                  \\
                            & KOH                   & ---                  \\ \midrule
Past MC                     & DM                    & $\checkmark$             \\ \midrule
Family MH                   & hyperlipidemia        & ---                  \\ \bottomrule
\end{tabular}
}
\end{table}

Results for In\_expB2 and TF-IDF retrieval models are presented in Table \ref{tab:extraction-results-other-models} of Appendix \ref{app:other_models}.
The In\_expB2 model on TREC CT 2021 data is the only one for which our query and document enrichment techniques are not improving results. 
We hypothesise that this is the case as the starting model (run 14) was already a very strong model compared to other baselines.
For the TF-IDF model, we can observe that the enrichment with current and past medical entities yields the best results both for 2021 and 2022 data.

Figure \ref{fig:relevant_2022_per_query} of Appendix \ref{app:topic_by_topic} presents a topic-by-topic analysis of the results in terms of the number of relevant trials ranked in top 20 using lexical models.
We can observe an incremental gain both from extracted entities and filtering.

\subsection{Effectiveness of filtering}

Next, we test several filtering methods as described in Section \ref{sec:method-filtering}.
As a base run, we take our best configuration from the previous experiment: BM25+ run enriching data with current medical conditions and medical history of the patient and family (run 14d).
Results for TREC CT 2021 are presented in Table \ref{tab:filtering-results}.

\begin{table*}[h]   
\centering  
\caption{   
Filtering results on TREC CT 2021 data. Letters describe the used filters: (A) Age, (G) Gender, (S) Smoking and (D) Drinking.  
\textbf{Bold} values indicate highest score overall.
Superscripts denote significant differences in paired Student's t-test with $p \le 0.05$. 
The identifier of each run is in the first column.
}  
\resizebox{0.98\textwidth}{!}{   
\begin{tabular}{c|l|ccccc}  
\toprule   
\textbf{\#}  
& \textbf{Model}   
& \textbf{nDCG@5}  & \textbf{nDCG@10}   & \textbf{P@10} & \textbf{RR}  & \textbf{\% filtered trials}\\
\midrule  
a &  
14. &  
0.508\hphantom{$^{bcdefg}$} &
0.462\hphantom{$^{bcdefg}$} &   0.276\hphantom{$^{bcdefg}$} &
  0.505\hphantom{$^{bcdefg}$} & --- \\
b &  
14d. &  
0.532\hphantom{$^{acdefg}$} &
0.480\hphantom{$^{acdefg}$} &   0.288\hphantom{$^{acdefg}$} &
  0.555\hphantom{$^{acdefg}$} & --- \\ \midrule
c &  
14d. + A &  
0.554$^{af}$\hphantom{$^{bdeg}$} &
0.509$^{abdf}$\hphantom{$^{eg}$} &   0.335$^{abdf}$\hphantom{$^{eg}$} &
  0.603$^{abdf}$\hphantom{$^{eg}$} & 23.4\% \\
d &  
14d. + G &   
0.537$^{f}$\hphantom{$^{abceg}$} &
0.483$^{f}$\hphantom{$^{abceg}$} &   0.288\hphantom{$^{abcefg}$} &
   0.556$^{b}$\hphantom{$^{acefg}$} & \ 5.7\% \\
e &  
14d. + AG &  
\textbf{0.561}$^{abdf}$\hphantom{$^{cg}$} &
\textbf{0.513}$^{abcdf}$\hphantom{$^{g}$} &   \textbf{0.337}$^{abdf}$\hphantom{$^{cg}$} &
  \textbf{0.604}$^{abdf}$\hphantom{$^{cg}$} & 26.3\% \\
f &  
14d. + SD &  
0.526\hphantom{$^{abcdeg}$} &
0.475\hphantom{$^{abcdeg}$} &   0.284\hphantom{$^{abcdeg}$} &
  0.546\hphantom{$^{abcdeg}$} & \ 0.7\% \\
g &  
14d. + AGSD & 
0.555$^{af}$\hphantom{$^{bcde}$} &
0.509$^{abdf}$\hphantom{$^{ce}$} &   0.335$^{abdf}$\hphantom{$^{ce}$} &
  0.595$^{abdf}$\hphantom{$^{ce}$} & 26.7\% \\
\bottomrule  
\end{tabular}  
}  
\label{tab:filtering-results}  
\end{table*}

Our filtering results align with other researchers' results, confirming that utilising age and gender fields can improve the quality of the final matching.
The usage of both filters (run $e$) removes, on average, 26.3\% trials from the top 1000 retrieved documents for all topics of the 2021 collection, improving the P@10 score by 4.9 percentage points over the unfiltered run.
Out of these two fields, the contribution of the age filter has more impact and is significantly better than the base run.

On the other hand, smoking and alcohol related-filtering does not help to improve the results further (runs $f$ and $g$). 
We grouped this filters together as our algorithm did not identify any smoker, and only nine drinking patients in TREC CT 2021 topics.
Despite only these few mentions, we observe deterioration of the results. 

\subsection{Neural re-ranking}
Table \ref{table:rerankers} shows the results of the re-ranking procedure discussed in Section \ref{sec:reranking}. We used the different models for re-ranking the results of a BM25 rank. We report the evaluations on the 2022 data. Models were trained on the 2021 data. Result of the TCRR model corresponds to the official TREC CT 2022 evaluation~\cite{roberts2022overview}.

\begin{table}[h]
    \centering
    \caption{Neural re-ranking evaluation results on TREC CT 2022 data. \underline{Underlined} values indicate highest score among general models. \textbf{Bold} values indicate highest score overall. $\dagger$-marked models indicate that there is a significant improvement over BM25 baseline.}
    \label{table:rerankers}
    \setlength\tabcolsep{2pt}
    \begin{tabular}{c|l|rrrr}
    \toprule
       \# &\textbf{Model} & \textbf{nDCG@5} & \textbf{nDCG@10} & \textbf{P@10} & \textbf{RR}\\
       \midrule
      & TREC median & --- & 0.392 & 0.258 & 0.411\\
    14  & BM25$_{all+inclusion}$ & 0.464 & 0.437 & 0.312 & 0.520\\  
    14d  & BM25$_{all+inclusion+cfp}$ & 0.502 & 0.460 & 0.328 & 0.521\\  \midrule
      & TopicalRR$_{}$ &0.558 &0.529 & 0.414 & \underline {0.630}\\
      & CriteriaRR$_{}$ & 0.382 & 0.387 & 0.294 & 0.428\\
      & TraditionalRR$_{}$ & 0.453 & 0.437 & 0.364 & 0.508\\
      & MonoBERT$_{}$ & 0.509 & 0.491 & 0.362 & 0.527\\
        \midrule
      &  TCRR$^\dagger$ &\underline {0.573} &\underline{0.557} & \underline{0.456}& 0.619\\ \midrule
       & TCRR$_{Bio}^\dagger$ & {0.627} & \textbf{0.604} & \textbf{0.482} & {0.672}\\
       & TCRR$_{Clinical}$ & 0.425 & 0.423 & 0.358 & 0.492\\
       & TCRR$_{Blue}$ & \textbf{0.631} & 0.583 & 0.452 & \textbf{0.691}\\

       \bottomrule
    \end{tabular}
\end{table}

As we hypothesise, in the context of CTs, the model benefits from the decomposition of the retrieval problem into two objectives, as it is experienced by TCRR (see Section \ref{sec:reranking}), the model exposed to the two learning objectives and best performing. We also provide results for TopicalRR and CriteriaRR, independently, which are the models exposed only to the first (topical relevance) and second (eligibility classification) learning objectives. Additionally, we present results for the regular re-ranking setup TraditionalRR.

For this set of experiments, we are mainly interested in the evaluation in terms of Precision since, in a real scenario, only eligible trials are considered. 
Given that on average other proposed systems perform poorly, as shown by the TREC CT median results~\cite{roberts2021overview,roberts2022overview}, precision (P@10) anywhere near 50$\%$ is regarded as a good result for this task. 
We analyse results from the proposed approach and find a significant improvement between the performance of TCRR models (TCRR and TCRR$_{Bio}$) and BM25  at a 95\% confidence level.
On average, this approach allows Bert-based models to gather more relevant documents than the selected baselines in the top 10.

We report results on different domain specific pre-trained models that we trained following our proposed approach. 
Again, we evaluated the best performing model, TCRR$_{Bio}$, in terms of Precision and found the improvement statistically significant.

\begin{figure}[ht]
    \centering
    \begin{subfigure}{1\textwidth}
    \includegraphics[width=\textwidth]{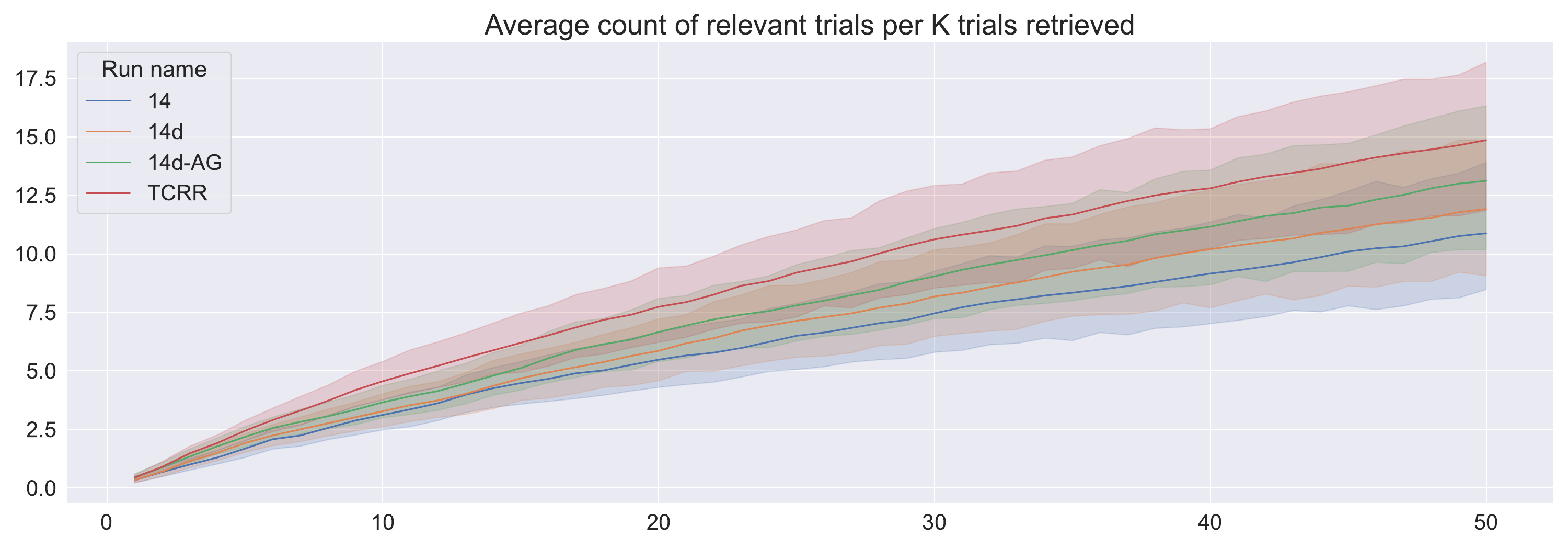}
\end{subfigure}
\hfill
\begin{subfigure}{1\textwidth}
    \includegraphics[width=\textwidth]{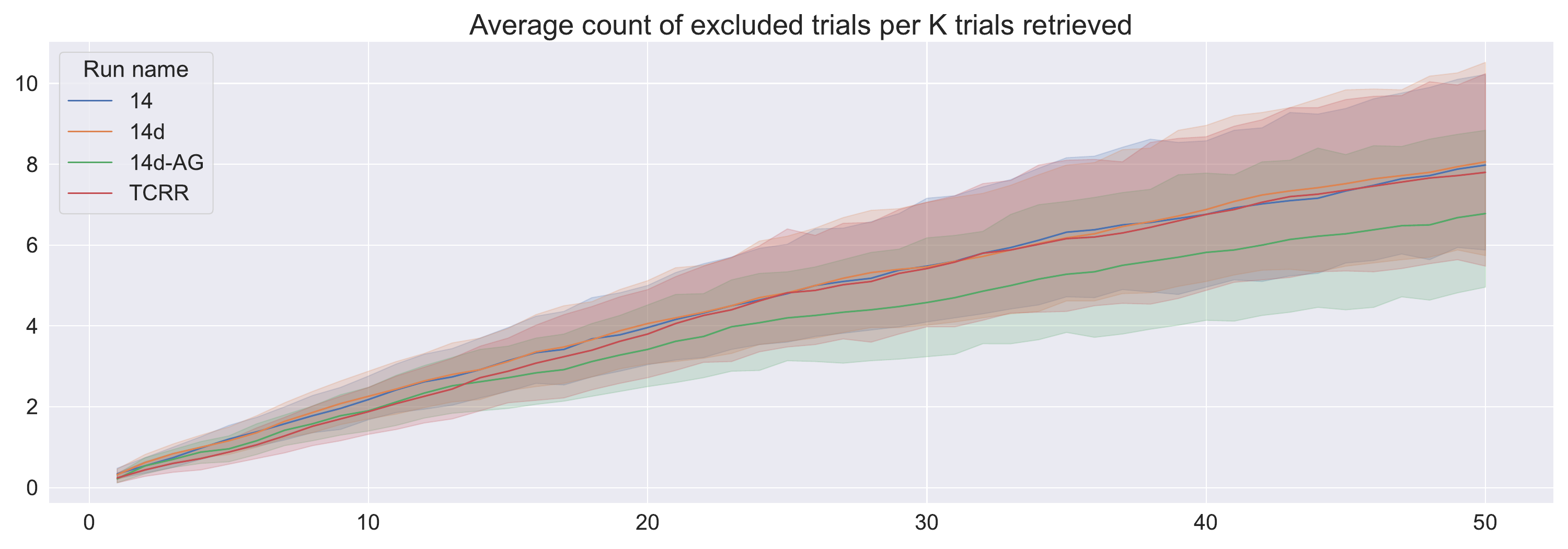}
\end{subfigure}
\caption{Averaged per patient count of relevant (top) and excluded (bottom) trials depending on a cut-off of K trials retrieved (x-axis) for TREC CT 2022 collection.}
\label{fig:averaged_count_2022}
\end{figure}

Figure \ref{fig:averaged_count_2022} presents two plots with an averaged per patient count of relevant and excluded trials depending on a cutoff point for the TREC CT 2022 collection.
Both techniques applied to lexical models, namely extracting drug and disease entities and filtering by age and gender, have a positive impact in finding more eligible trials.
However, only the run with filtering is able to retrieve consistently fewer ineligible trials than the baseline run.
We can also see that, on average, our best non-neural run (14d-AG), retrieves twice as many trials for which a patient is eligible than excluded.
Similarly, the TCRR neural re-ranking is further improving the number of relevant trials, but helps in removing ineligible only for the first 15 trials.
One possible explanation is that we re-ranked only the top 50 trials retrieved by the first-stage ranker.

\section{Discussion}

In this work, we revisit the pipeline-based model for patient-to-CT matching. 
First, we report an extensive set of experiments for the first stage retrieval and propose an effective enrichment procedure to get the best out of the initial ranks. 
Second, we propose an adaptation of training a cross-encoder to the CT problem, taking advantage of the structured nature of the considered documents and the task.

We find that the inclusion criteria section has the most considerable impact on the retrieval score for all three tested lexical models meaning that these models cannot use all the available information.
These outcomes motivate our further work in structuring queries and documents using entity extraction and negation classification methods.
The results show improvements in finding relevant trials when applying data enrichment methods.

We show results for experiments on different configurations of our pipeline and compare our approach with different models previously used for the task. 
We focus on BERT-based models, which so far have not necessarily outperformed probabilistic lexical ranking models for the clinical trial matching task. Even though the results in Table~\ref{table:rerankers} also show how changing the initial weights of the model can affect the overall performance (i.e., by choosing domain-specific model like BioBERT), we show that the improvements of our proposed approach are not due to the selection of a domain-specific pre-trained model, which is the case of the TCRR. 

These results also provide an idea of which pre-trained model fits the task best. Overall, the TCRR initialised with BioBERT weights shows promising results, while ClinicalBERT weights were not the best choice in this scenario.

To our knowledge, this study is the first to focus on enriching documents and queries showing gains in the models' ability to find more eligible trials. 
Furthermore, our novel re-ranking concerning eligibility shows additional improvement for this task, comparable to the more expensive approach using the T5 architecture~\cite{pradeep2022neural}. 

Our proposed re-ranking formula is different as it explicitly models the eligibility decisions instead of using only the topical relevance.
This distinguishes our study from the previous works concerning clinical trial re-ranking~\cite{rybinski2020clinical}. 

Although this work focused on CT retrieval, we believe the approach can also be applied to other IR tasks where first, they involve ranking documents based on topics, and, in a second instance, the retrieval results are tailored by considering more specific criteria or constraints. One example of such a task is the selection of primary studies (citation screening) for the systematic literature reviews~\cite{Kusa2022AutomationStudy}.

There are several limitations of this study, both related to the dataset and the models. 
Usage of the TREC CT collection implies that the patient descriptions are relatively short, i.e., EHR admission note-style documents.
We acknowledge that our approaches could have problems handling longer sequences.

Additional limitations are related to the amount of data available for training and evaluating systems on the CT retrieval task. This issue, in our study, explicitly affects the curriculum learning scenario in the eligibility determination objective. It may limit the model in learning relevant patterns needed to scale to different clinical settings or patient populations.

Furthermore, the topics are written only in English.
This does not concern clinical trials, for which the ClinicalTrials.gov database is the leading international source.
Nevertheless, multilingual medical retrieval may present challenges for both lexical and neural models, as the nuances and complexities of medical terminology can vary significantly across languages.
Addressing these limitations and developing strategies for multilingual medical retrieval is an essential area for future research.

\section{Conclusion}

This paper presents an approach for clinical trial retrieval under the patient-to-trial paradigm.
We investigate the impact of individual clinical trial sections showing that the `inclusion' section alone contributes the most to the final retrieval score.
Moreover, we evaluate the handling of complex eligibility criteria for matching patients to clinical trials by combining input from information extraction modules into a lexical retrieval model.
The extracted drug and disease entities and their negations positively impact the retrieval of eligible trials.
On the other hand, filtering based on gender and age proved to be successful in eliminating ineligible trials.

Additionally, we propose an effective training strategy for neural re-ranking of clinical trials based on two distinct learning objectives. 
The first objective is the traditional relevance objective, while the second objective focuses on giving importance to the eligibility criteria and involves a classification objective that distinguishes between eligible and discarded samples.
Our results indicate that even with limited data, this model is capable of further improving the Precision of our approach.
Even though our proposed system involves many single components, it showcases an alternative approach to the clinical trial matching problem, emphasising the importance of eligibility criteria.
In future work, we plan to measure the impact of extracted entities on neural re-ranking models. 

\begin{ack}
This work was supported by the EU Horizon 2020 ITN/ETN on Domain Specific Systems for Information Extraction and Retrieval -- DoSSIER (H2020-EU.1.3.1., ID: 860721).
\end{ack}

\appendix

{
\small
\bibliography{references}
\bibliographystyle{plainnat}
}

\newpage

\clearpage

\section{Datasets summary} \label{app:datasets}

A summary of datasets is presented in Table \ref{tab:trec-ct-stats}.

\begin{table}[h]
\centering
\caption{Statistics of TREC CT datasets from 2021 and 2022.}
\label{tab:trec-ct-stats}
\begin{tabular}{@{}lrr@{}}
\toprule
                 & {2021} & {2022} \\ \midrule
Documents        & 375,580                          & 375,580                          \\
Topics (patient notes)           & 75                               & 50                               \\
Avg. topic length (tokens)           & 133.4                               & 105.9                               \\
Avg. topic length (sentences)           & 11.2                               & 9.4                               \\ \midrule
Total judgements & 35,832                           & 35,394                           \\ 
\hspace{1cm}-- Eligible (2)        & 5,570                            & 3,939                            \\
\hspace{1cm}-- Excluded (1)         & 6,019                            & 3,036                            \\
\hspace{1cm}-- Not relevant (0)    & 24,243                           & 28,419                           \\ 
\midrule
Unique trials judged   & 26,162  & 26,585 \\
\bottomrule
\end{tabular}
\end{table}

\section{Other lexical models} \label{app:other_models}

Table \ref{tab:input_fields_other_models} presents results for the clinical trial documents sections impact on the ranking with In\_expB2 and TF-IDF models.

Table \ref{tab:extraction-results-other-models} shows results for the query and document enrichment experiment with In\_expB2 and TF-IDF models.

\begin{table}[!h]
\centering
\caption{Impact of CT sections on the performance of In\_expB2 and TF-IDF retrieval models. 
For each model, the first group contains results using only a single section as a document representation, and the second group represents results using several concatenated sections.
`Criteria' contains the value from the eligibility section, whereas `inclusion' and `exclusion' contains only the list of heuristically split section. 
{\ul Underlined} values indicate highest score within the group, \textbf{bold} values indicate highest score overall for each model.
The identifier of each run is in the first column.}
\label{tab:input_fields_other_models}
\resizebox{\textwidth}{!}{%
\begin{tabular}{@{}l|l|cccc|cccc@{}}
\toprule
    &                                                 & \multicolumn{4}{c|}{\textbf{TREC CT 2021}}                                                     & \multicolumn{4}{c}{\textbf{TREC CT 2022}}                                                      \\ \midrule
    \textbf{\#}  
& \textbf{Input sections}                                  & \textbf{nDCG@5}  & \textbf{nDCG@10}   & \textbf{P@10} & \textbf{RR}                  & \textbf{nDCG@5}  & \textbf{nDCG@10}   & \textbf{P@10} & \textbf{RR}                  \\ \midrule
&  \multicolumn{4}{l}{\textbf{In\_expB2}}  \\ \midrule
1.  & brief title                                    & .174                & .167                & .112                & .221                & .222                & .193                & .134                & .296                \\
2.  & official title                                 & .194                & .179                & .109                & .252                & .245                & .209                & .146                & .380                \\
3.  & description (\emph{desc.})                    & .354                & .327                & .200                & {\ul.458}                & .341                & .296                & .186                & .380                \\
4.  & summary (\emph{sum.})                         & .299                & .288                & .172                & .313                & .373                & .322                & .222                & {\ul .511}          \\
5.  & conditions  (\emph{cond.})                   & .119                & .119                & .081                & .172                & .141                & .135                & .094                & .196                \\
6.  & inclusion                                       & {\ul.398}           & {\ul.370}           & {\ul.225}           & .445           & {\ul .389}          & {\ul.345}                & {\ul .238}          & .485                \\
7.  & exclusion                                       & .147                & .131                & .051                & .134                & .138                & .133                & .070                & .142                \\
8.  & criteria                                      & .386                & .360                & .192                & .409                & .347                &  .322          & .224                & .409                \\ \midrule
9.  & brief title + official title (\emph{tit.})  & .252                & .235                & .149                & .325                & .300                & .248                & .162                & .423                \\
10. & sum. + criteria + tit.                          & .454                & .426                & .227                & .467                & .474                & .435                & .286                & {\ul \textbf{.574}} \\
11. & desc. + criteria + tit.                         & .462                & .437                & .248                & .419                & .437                & .417                & .292                & .441                \\
12. & sum. + desc. + tit.                             & .441                & .405                & .252                & .517                & .455                & .415                & .282                & .488                \\
13. & sum. + desc. + tit. + cond. (\emph{all})      & .440                & .411                & .252                & .533                & .463                & .420                & .282                & .493                \\ 
14. & all + inclusion                                 & {\ul \textbf{.518}} & {\ul \textbf{.482}}                & {\ul \textbf{.281}}                & {\ul \textbf{.553}} & {\ul \textbf{.506}}                & {\ul \textbf{.480}} & {\ul \textbf{.346}} & .539                \\
15. & all + exclusion                                 & .395                & .365                & .203                & .377                & .425                & .388                & .254                & .473                \\
16. & all + criteria                                  & .480                & .455 & .267                & .441                & .490 & .449                & .312                & .508                \\ \midrule \midrule
&  \multicolumn{4}{l}{\textbf{TF-IDF}}  \\ \midrule
1.  & brief title                                    & .196                & .172                & .107                & .253                & .221                & .193                & .130                & .305                \\
2.  & official title                                 & .203                & .181                & .109                & .256                & .238                & .200                & .138                & .353                \\
3.  & description (\emph{desc.})                    & .313                & .280                & .160                & .396                & .309                & .272                & .162                & .387                \\
4.  & summary (\emph{sum.})                         & .281                & .263                & .147                & .327                & .336                & .288                & .196                & {\ul .496}          \\
5.  & conditions  (\emph{cond.})                   & .124                & .127                & .087                & .180                & .152                & .144                & .094                & .201                \\
6.  & inclusion                                       & {\ul.411}           & {\ul.377}           & {\ul.229}           & {\ul.466}           & {\ul .383}          & {\ul.333}                & {\ul .232}          & .444                \\
7.  & exclusion                                       & .145                & .132                & .053                & .146                & .139                & .129                & .072                & .125                \\
8.  & criteria                                      & .383                & .364                & .199                & .421                & .338                &  .316          & .220                & .405                \\ \midrule
9.  & brief title + official title (\emph{tit.})  & .235                & .213                & .129                & .300                & .276                & .223                & .146                & .397                \\
10. & sum. + criteria + tit.                          & .444                & .411                & .214                & .436                & .416                & .389                & .260                & .497 \\
11. & desc. + criteria + tit.                         & .458                & .429                & .232                & .435                & .403                & .385                & .264                & .438                \\
12. & sum. + desc. + tit.                             & .364                & .335                & .195                & .429                & .392                & .354                & .236                & .480                \\
13. & sum. + desc. + tit. + cond. (\emph{all})      & .362                & .332                & .184                & .435                & .405                & .358                & .234                & {\ul \textbf{.505}}                \\ 
14. & all + inclusion                                 & .478 & .446               & {\ul \textbf{.260}}                & {\ul \textbf{.481}} & .430                & .406 & {\ul \textbf{.282}} & .474                \\
15. & all + exclusion                                 & .380                & .345                & .183                & .381                & .380                & .342                & .222                & .454                \\
16. & all + criteria                                  & {\ul \textbf{.482}}                & {\ul \textbf{.450}} & .248                & .454                & {\ul \textbf{.437}} & {\ul \textbf{.407}}                & .274                & .478                \\ \bottomrule

\end{tabular}%
}
\end{table}

\begin{table}[!h]
\centering
\caption{Experimental results for runs with index and query expanded with extracted entities for In\_expB2 and TF-IDF retrieval models. 
Letters describe usage of extracted affirmative and negative medical entities for (c) current conditions, (p) past conditions, and (f) family history. 
\textbf{Bold} values indicate highest score overall for each model.
The identifier of each run is in the first column.
}
\label{tab:extraction-results-other-models}
\resizebox{\textwidth}{!}{%
\begin{tabular}{@{}l|r|cccc|cccc@{}}
\toprule
     & \multicolumn{1}{l|}{}      & \multicolumn{4}{c|}{\textbf{TREC CT 2021}}                               & \multicolumn{4}{c}{\textbf{TREC CT 2022}}                              \\ \midrule
     \textbf{\#}  
& \textbf{Model}    & \textbf{nDCG@5}  & \textbf{nDCG@10}   & \textbf{P@10} & \textbf{RR}             & \textbf{nDCG@5}  & \textbf{nDCG@10}   & \textbf{P@10} & \textbf{RR}            \\ \midrule
&  \multicolumn{4}{l}{\textbf{In\_expB2}}  \\ \midrule
14.  & all + inclusion            & \textbf{.518} & \textbf{.482}                & \textbf{.281}                & \textbf{.553} & .506                & .480 & .346 & .539        \\ 
16.  & all +\ \ \  criteria       & .480          & .455           & .267          & .441          & .490           & .449          & .312           & .508          \\ \midrule
14a. & + c\ \ \ \ \               & .499          & .457           & .272          & .483           & .515          & \textbf{.484}           & .340         & .555          \\
14b. & + cf\ \ \ \                & .491          & .455           & .272          & .479           & \textbf{.524}           & .482          & \textbf{.342}          & .554 \\
14c. & + cp\ \ \                  & .494          & .461           & .267           & .494           & .521          & .479          & .336          & \textbf{.559}          \\
14d. & + cfp\ \                   & .492          & .457           & .267            & .490           & .521          & .475          & .332          & .547          \\ \midrule \midrule
&  \multicolumn{4}{l}{\textbf{TF-IDF}}  \\ \midrule
14.  & all + inclusion            & .478 & .446                & .260                & .481 & .430                & .406 & .282 & .474           \\ 
16.  & all +\ \ \  criteria       & .482           & .450        & .248                & .454                & .437 & .407                & .274                & .478          \\ \midrule
14a. & + c\ \ \ \ \               & \textbf{.484}          & .452           & .259          & .463           & \textbf{.496} & \textbf{.439} & .302 & \textbf{.536}          \\
14b. & + cf\ \ \ \                & .481          & .446           & .259          & .457           & .493           & .437          & .302          & .528 \\
14c. & + cp\ \ \                  & .483          & \textbf{.459}           & \textbf{.261} & \textbf{.515}           & .475          & \textbf{.439}          & \textbf{.306}          & .511          \\
14d. & + cfp\ \                   & .477          & .453 & .259 & .508 & .477          & \textbf{.439}          & .304          & .517          \\ \bottomrule
\end{tabular}%
}
\end{table}

\newpage

\section{Topic-by-topic analysis}  \label{app:topic_by_topic}

\begin{figure}[ht]
    \centering
    \begin{subfigure}{1\textwidth}
    \includegraphics[width=\textwidth]{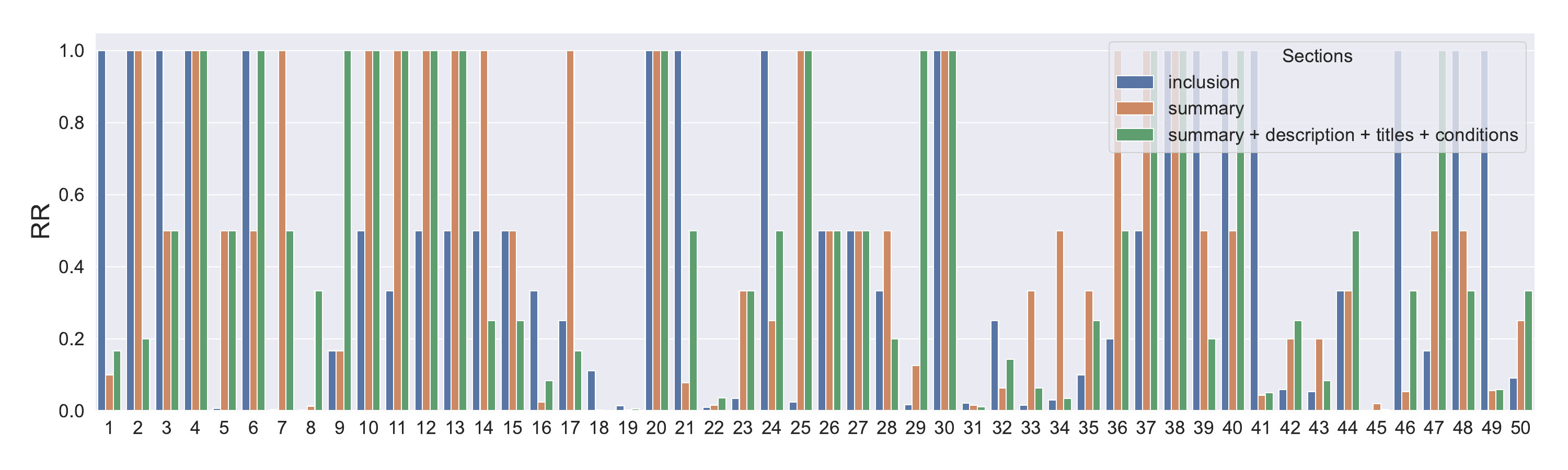}
\end{subfigure}
\hfill
\begin{subfigure}{1\textwidth}
    \includegraphics[width=\textwidth]{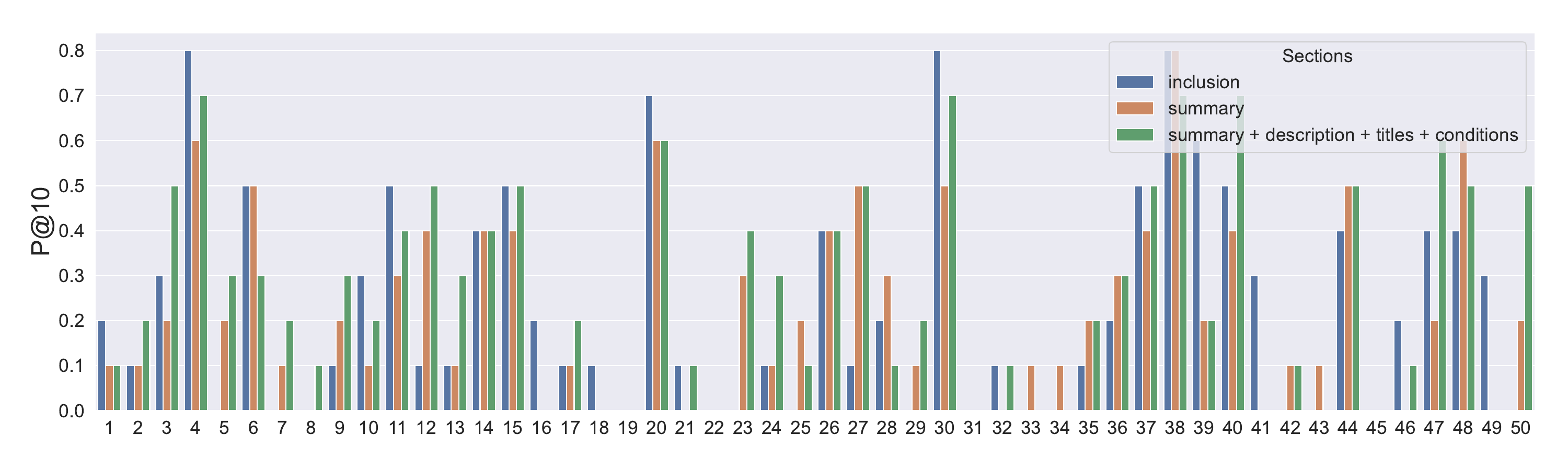}
    \label{fig:second}
\end{subfigure}
\caption{Reciprocal Rank (top) and P@10 (bottom) comparison for a BM25+ model with different document representations for TREC CT 2022 data.}
\label{fig:sections_2022_per_query}
\end{figure}

Figure \ref{fig:sections_2022_per_query} shows topic-by-topic comparison for RR and P@10 for BM25+ using \emph{inclusion} (run 6) \emph{summary} (run 4) and \emph{summary}, \emph{description}, \emph{titles} and \emph{condition} sections concatenated (run 13).

\begin{figure}[h]
    \centering
\includegraphics[width=\textwidth]{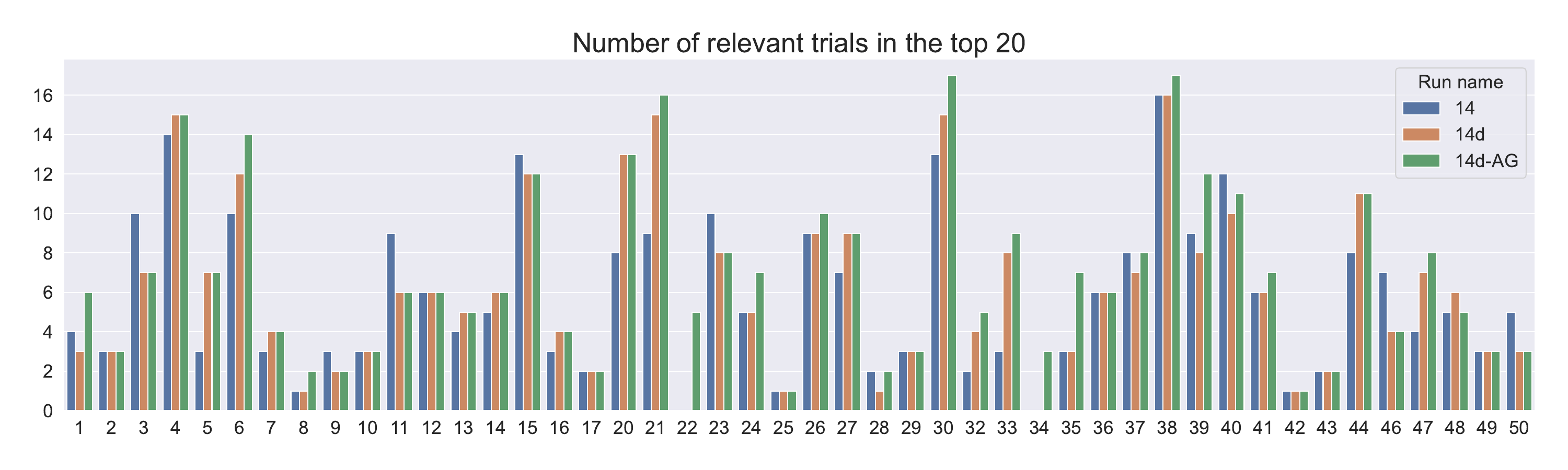}
\caption{Number of relevant trials in the top 20 for the three best BM25+ runs from each experiment: 14 -- baseline, 14d -- further query and index enriched with extracted entities, and 14d+AG -- further filtered for age and gender.}
\label{fig:relevant_2022_per_query}
\end{figure}

Figure \ref{fig:relevant_2022_per_query} presents the number of relevant trials at the top 20 retrieved trials for the three best BM25+ runs from each experiment.

\end{document}